\documentclass[prfluids,showpacs,notitlepage,superscriptaddress,longbibliography]{revtex4-2}

\usepackage{epsf,graphicx,amssymb}
\usepackage[usenames]{color}
\usepackage{natbib}
\usepackage{float,ulem}
\usepackage{gensymb}
\usepackage{amsmath,amssymb}
\usepackage{color,soul}
\usepackage{lipsum} 
\usepackage{physics}
\usepackage[exponent-product=.,inter-unit-product=\,, binary-units]{siunitx}
\usepackage{todonotes}
\usepackage{xcolor}
\usepackage{enumitem}
\usepackage{interval}

\usepackage{hyperref}

\hypersetup{
    colorlinks=true,
    linkcolor=blue,
    citecolor=magenta,
    urlcolor=blue,
    breaklinks=true, 
    bookmarksopen=true, 
    }
\newcommand{\veps}{\varepsilon}
\newcommand{\vphi}{\varphi}
\renewcommand{\emph}[1]{\textit{#1}}
\newcommand{\emax}{\eta_\text{max}}
\newcommand{\esat}{\eta_\text{sat}}
\newcommand{\emin}{\eta_\text{min}}
\usepackage[percent]{overpic}

\begin{document}

\title{Axisymmetric gravity-capillary standing waves on the surface of a fluid}
\author{Jules Fillette}
\affiliation{Laboratoire de Physique de l'École Normale Supérieure, CNRS, PSL Research University, Sorbonne Université, Université Paris Cité, F-75005 Paris, France}
\affiliation{Universit\'e Paris Cité, CNRS, MSC, UMR 7057, F-75013 Paris, France}
\author{Stéphan Fauve}
\affiliation{Laboratoire de Physique de l'École Normale Supérieure, CNRS, PSL Research University, Sorbonne Université, Université Paris Cité, F-75005 Paris, France}
\author{Eric Falcon}
\email[E-mail: ]{eric.falcon@u-paris.fr (corresponding author)}
\affiliation{Universit\'e Paris Cité, CNRS, MSC, UMR 7057, F-75013 Paris, France}


\begin{abstract}We report on the experimental study of axisymmetric gravity-capillary standing waves generated by a vertically vibrating ring partially immersed into a fluid. Different regimes of standing waves are highlighted at the basin center depending on the forcing parameters: linear, nonlinear and ejection regimes. For weak forcing, the standing waves display a resonant response, close to a natural frequency of the circular basin, predicted by the linear theory. For stronger forcing, we observed that the experimental spatial profile of standing waves breaks the up-down symmetry, and is well described by a third-order nonlinear theory. When the forcing is further increased, the maximum height of the axisymmetric wave crest at the basin center is found to increase linearly with its wavelength, due to the saturation of its steepness, a result well captured by a proposed model.
\end{abstract}

\maketitle


\section{Introduction}
One of the most common wave observations in everyday life is the propagation of concentric waves after a stone has disturbed the interface between water and air~\cite{LeMehaute}. Looking at this pattern, one could wonder what happens when the waves converge instead of diverging. This phenomenon, characterized by the concentration of a finite amount of energy in an infinitely small area, is called wave focusing. Wave focusing has been studied in optics since the 19th century, in the neighborhood of a caustic~\cite{Airy}, or a Huygens cusp of light~\cite{LightCusp}. In particular, diffraction theory states that (in a homogeneous medium with no source) the diffraction limit, i.e., the shortest spatial wave-field fluctuations, is precisely one-half wavelength, $\lambda/2$~\cite{PcpOfOptics} and focusing is known to shift the phase of the wave~\cite{GouyPhase}. In acoustics, wave focusing has been used to develop tools for trapping or tweezers~\cite{AcousticTrapping, AcousticTwizzers}, whereas time-reversal techniques overcome the diffraction limit and reduce the size of the focal spot as narrow as $\lambda/14$~\cite{RosnyPRL2002}. Although hydrodynamic systems have several advantages compared to optics or acoustics (macroscopic, slow dynamics, and direct space-and-time resolved wave-field measurement), hydrodynamic focusing has not been studied in detail except for spatial focusing with a parabolic shaped wave maker~\cite{FocusChavarria2018}, wave control by time reversal and holography methods~\cite{BacotNP2016} or three-dimensional wave breakings~\cite{SheAOR1997}. Nevertheless, directional focusing has also been suggested as a candidate for the formation of rogue waves~\cite{Fochesato2007} and amplification of tsunamis~\cite{BerryPRSA2007} in the ocean.

Axisymmetric surface waves have been routinely studied in the past. Indeed, the behavior of standing waves in a circular basin is of primary interest, in particular to the study of sloshing in cylindrical tanks or harbor oscillations \cite{Ibrahim2005,MichelPRF2017}. Experiments in large-scale basins were also reported in which converging axisymmetric gravity waves are generated by several wavemakers, driven in unison, surrounding the tank~\cite{Minoura2011,Maeda2004,McAllister-JFM2022}. These studies mainly focus on the transient phenomenon of jetting occurring at the center of the tank. Such hyperbolic-shape jet eruptions on a fluid surface have also been investigated, within more feasible setups, either by drop or projectile impact~\cite{Worthington1882,Josserand2017,Che2018,TruscottARFM2014}, bubble bursting at a free surface~\cite{GhabachePoF2014} or parametric forcing (Faraday instability) of cylindrical containers~\cite{Longuet1983,Zeff2000}. These observations are usually compared to gravity-wave profiles from linear~\cite{Lamb-Book} or nonlinear~\cite{Mack1962,TsaiPoF1987,Dalzell1999} theories. Recently, numerical simulations investigated the decay of axisymmetric gravity-capillary waves initially generated by a zeroth-order Bessel-function deformation~\cite{FarsoiyaJFM2017,Basak-JFM-2020}. All these experimental studies mainly concern the transient regimes, and axisymmetric gravity-capillary \textit{standing} waves have been much less experimentally investigated.

Here, we propose an original model experiment to study axisymmetric gravity-capillary standing waves generated by a vertically vibrating ring on a fluid surface. Under weak sinusoidal forcing, the spatial pattern of the waves is found to agree with linear predictions~\cite{Lamb-Book}. For high enough forcing, the up-and-down symmetry of the spatial profile is broken as predicted by a nonlinear theory~\cite{TsaiPoF1987,Mack1962}, and a divergence of the wave amplitude occurs at the basin center, sometimes with the ejection of a drop. We show in particular that the maximum height reached by the axisymmetric gravity-wave crest, at the basin center, increases linearly with its wavelength, due to the saturation of its steepness to $5/(2\pi)$. To the best of our knowledge, this saturation has not been previously reported. It should not be confused with the stability limit of a periodic sharp-crested wave derived for unidimensional~\cite{Stokes1847,Penney1952, Grant1973} or axisymmetric~\cite{Mack1962} standing waves of finite amplitude, and tested experimentally~\cite{Taylor1953,Longuet1983,Fultz1963,McAllister-JFM2022}.
\begin{figure}[!htb]
	\centering
	\includegraphics[width=0.5\linewidth]{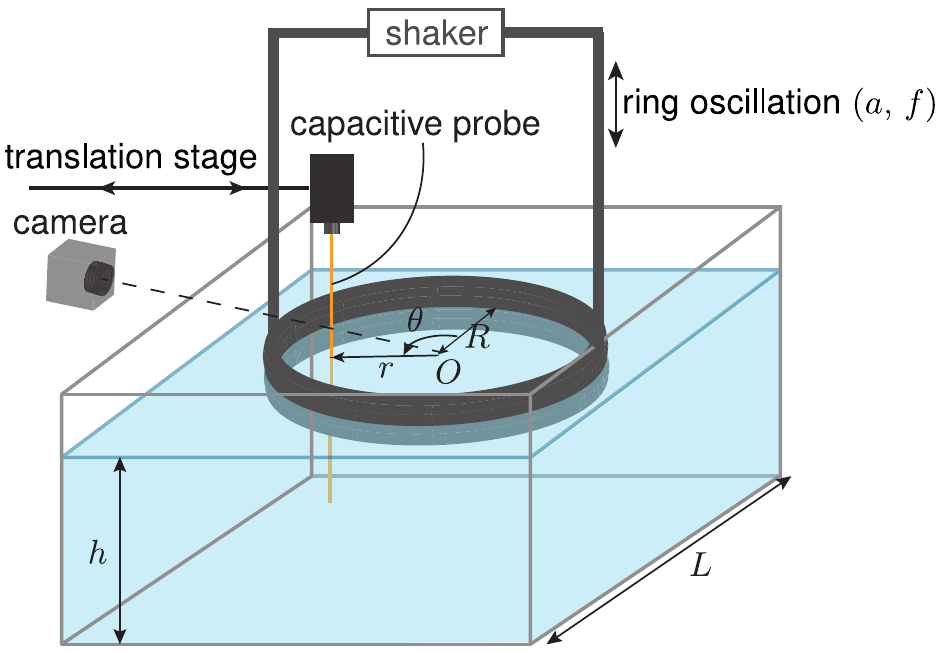}
	\caption{Experimental setup of axisymmetric gravity-capillary wave focusing. The surface elevation $\eta(r,t)$ is measured using a capacitive probe mounted on a translation stage, and a side camera.} 
	\label{fig:Dispositif}
\end{figure}

\section{Experimental setup}
The experiment consists of a cubic transparent tank ($L = \SI{19}{\centi\meter}$ wide) filled with distilled water (density $\rho = \SI{1000}{\kilogram\per\cubic\meter}$) up to a depth $h = \SI{7}{\centi\meter}$ (see Fig.~\ref{fig:Dispositif}). We add surfactants to fix its surface tension $\gamma$ to a constant value of $\SI{37}{\milli\newton\per\meter}$ by using Trimethyl(tetradecyl)ammonium bromide at a concentration higher than the critical micelle concentration~\cite{Barry-JCIS1970, More-Springer2015}. Axisymmetric convergent waves are generated by the vertical oscillations of a solid ring made of plexiglass (internal radius $R = \SI{8.25}{\centi\meter}$, vertical thickness \SI{2}{\centi\meter}) half immersed into the fluid at rest. The ring is mechanically connected to an electromagnetic shaker (Dynamic Solution VTS-100) driven by a sinusoidal voltage from a power supply (Kepco 36V/6A) leading to a vertical ring motion $z(t) = \flatfrac{a \sin(2 \pi f t)}{2}$ where $f$ and $a$ are the forcing frequency and amplitude, respectively ($f\in[5,9.3]$~Hz, i.e., $\lambda \in[2.3,6.5]$~cm and $a\in[0,1]$~cm). A point of the free surface is referred to by its polar coordinates $(O,r,\theta)$. Nevertheless, $\theta$ is not considered hereafter as the phenomenon is mainly axisymmetric. Indeed, special attention is paid to adjusting the horizontality of the ring, and limiting the transverse vibrations, that break this symmetry. After a few forcing periods, the transient regime vanishes. The surface elevation $\eta(r,t)$ of the stationary wave field is then measured at a single location over time $t$, thanks to a home-made capacitive probe (\SI{10}{\micro\meter} in vertical resolution)~\cite{PRL2007-fil}. We iterate this temporal measurement for every $r$ along the ring diameter (with a \SI{1}{\milli\meter} step) using a translation stage with a stepper motor driven by a computer. Moving the probe along a diameter, therefore, gives access to the wave profile resolved in time and the wave envelope resolved both in space and time. The corresponding vertical resolution is \SI{100}{\micro\meter}. The nonlinear parameter, namely the wave steepness, is $\veps \equiv \flatfrac{\emax}{\lambda}$ where $\lambda$ is the wavelength and $\emax$ is the maximum elevation at the container center $r=0$. $\veps$ is varied by almost two decades in the range $\veps \in[0.01,1]$.
\begin{figure}[!htb]
	\centering
	\includegraphics[width=1\linewidth]{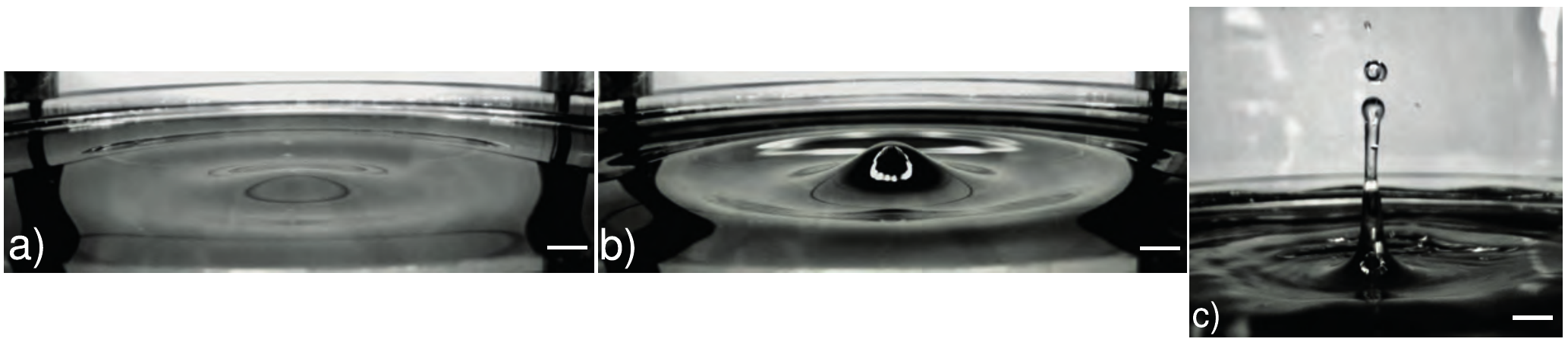}
	\caption{Typical regimes of axisymmetric standing waves for increasing forcing amplitudes (side views). (a) Linear regime ($\veps \simeq 0.07$, $a = \SI{0.04}{\centi\meter}$). (b) Nonlinear regime ($\veps \simeq 0.2$, $a = \SI{0.18}{\centi\meter}$). (c) Ejection regime ($\veps \simeq 0.7$, $a = \SI{0.36}{\centi\meter}$). The white bar corresponds to \SI{1}{\centi\meter}. Sinusoidal forcing $f = \SI{6.75}{\hertz}$. Part of the transparent solid ring is visible at the back.} 
	\label{fig:Regimes}
\end{figure}

\section{Patterns}\label{regimes}
Different typical axisymmetric patterns of the free surface are observed depending on the control parameters $a$ and $f$. We show in Fig.~\ref{fig:Regimes} the qualitative influence of increasing the forcing amplitude $a$ (from left to right), for a fixed forcing frequency $f$ (see also movies in Supplemental Material \cite{SuppMat}). At low $a$ ($\veps \simeq 0.07$), standing axisymmetric oscillations are gentle in particular near the center (see Fig.~\ref{fig:Regimes}a). We call below this regime the \emph{linear regime}. For high enough $a$ ($\veps > 0.1$), nonlinearities arise and the up-and-down central deformation is more prominent and much higher than the periphery ones (see Fig.~\ref{fig:Regimes}b). This regime is called afterwards the \emph{nonlinear regime}. When the forcing amplitude $a$ is further increased ($\veps \gtrsim 0.7$), we observe an \emph{ejection regime} characterized by the formation of a thin and intense jet at the center, with the ejection of at least one droplet.

\begin{figure}[!htb]
	\centering
	\includegraphics[trim=0.2cm 0cm 1.0cm 0.6cm, clip=true,width = 0.6\linewidth, height = 0.45\linewidth]{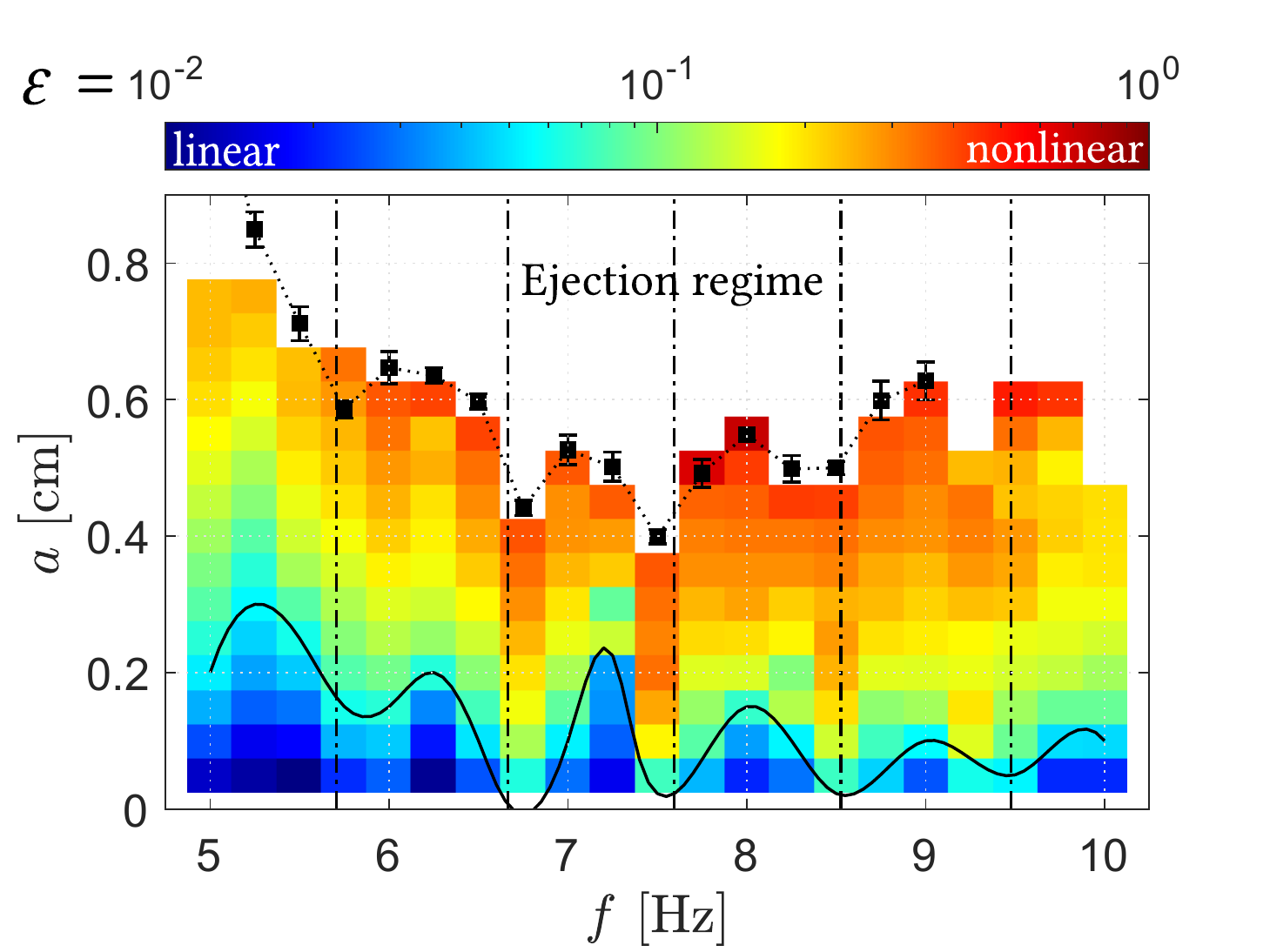}
	\caption{Wave steepness, $\veps= \flatfrac{\eta_\text{max}}{\lambda}$, of the central deformation as a function of the forcing amplitude $a$ and frequency $f$. Logscale colorbar. Vertical lines: circular basin eigenfrequencies $f_n$ from $J_0'(k_nR) = 0$ (see text) where $f_n$ and $k_n$ are related by Eq.~\eqref{eq:disprelation}. Solid line corresponds to the same value of $\veps \simeq \num{0.1}$ as a function of $f$. ($\blacksquare$): maximum amplitude of the central deformation before ejection as a function of $f$. } 
	\label{fig:PhaseDiagram}
\end{figure}

\section{Phase diagram}\label{phase}
We now explore in more detail the phase diagram of the three regimes found in \S \ref{regimes} as a function of the control parameters. We report in Fig.~\ref{fig:PhaseDiagram} the measured values of the wave steepness $\veps$ for each accessible pair ($f$, $a$) of the forcing parameters. Following a virtually vertical line (from bottom to top) in Fig.~\ref{fig:PhaseDiagram}, $\veps$ is found to increase with $a$ from very weak values ($\simeq \num{e-2}$ -- in blue) to values close to unity (in red). The ejection regime occurs when crossing the black-dotted line. The influence of the forcing frequency is highlighted by following the black solid line, which links data with the same value of $\veps$ (namely $\veps \simeq 0.1$). The curve minima point out frequencies for which the central deformation reaches this specific steepness although the forcing amplitude is weak, and thus correspond to resonances of the system. Moreover, these resonance frequencies appear to be in good agreement with the main theoretical eigenmodes of the ring marked by vertical dash-dotted lines. Indeed, the axisymmetric eigenmodes of a circular basin are obtained by considering an inviscid, irrotational, and incompressible fluid whose velocity potential $\vphi$ satisfies the Laplace equation $\Delta \vphi = 0$. The solution implies the Bessel function of the first kind of order $\alpha$, $J_{\alpha}(kr)$, where $k$ is the wave vector $k = \flatfrac{2\pi}{\lambda}$ \cite{note}. Moreover, imposing that the fluid cannot penetrate the solid boundary at $r = R$ over $t$ leads, in the linear approximation, to $\left. J_0'(x) \right|_{k_nR} = 0$, where the prime stands for the spatial derivative of $J_0(x)$~\cite{Lamb-Book}. This quantifies the modes of the system to discrete wave vectors~$k_n$. Using the linear dispersion relation of inviscid deep-water gravity-capillary waves~\cite{Lamb-Book} (as $kh > 8$ for $\lambda < \SI{5}{\centi\meter}$),
\begin{equation}
	\omega = \sqrt{gk + \frac{\gamma}{\rho} k^3} \, ,
	\label{eq:disprelation}
\end{equation}
the corresponding axisymmetric eigenfrequencies $f_n$ read $5.70$, $6.64$, $7.55$, $8.47$, and $9.39$~\si{\hertz} for $n = 3$, $4$, $5$, $6$, and $7$, respectively ($g = \SI{9.81}{\meter\per\square\second}$ is the acceleration of gravity). Note that the linearized kinematic condition at the interface $z=\eta(r,t)$ leads to $\pdv*{\vphi}{z} = \pdv*{\eta}{t}$, where $\vphi$ is the velocity potential, and implies that $\vphi$ and $\eta$ have the same dependence on $r$. Moreover, experiments show that the fundamental angular frequency $\omega$ of waves coincides with the forcing pulsation, $2\pi f$, leading thus to the same notation. Note that the vertical thickness of the ring is finite and the no-penetration condition is not fully verified below the ring (see \S \ref{Profile}). However, we have verified that the initial immersed depth of the ring into the water does not impact qualitatively the wave shape (see Appendix~\ref{AppendA}), but could explain slight departures between the resonances and eigenfrequencies in Fig.~\ref{fig:PhaseDiagram}. We have also verified, using a ring of smaller diameter, that the resonances correspond to the eigenfrequencies, although the latter change (see Appendix~\ref{AppendB}). The ejection threshold is hardly visible in Fig.~\ref{fig:PhaseDiagram} outside the range $f \in [5, 9]$\,Hz mainly because, for smaller $f$, the wavelength becomes comparable to the system size [e.g., $\lambda|_{\SI{5}{\hertz}} = \SI{6.5}{\centi\meter}$ from Eq.~\eqref{eq:disprelation}], so that the standing wave pattern is modified by finite size effects of the container. For $f> \SI{9}{\hertz}$, the ejection regime cannot be reached due to the mechanical limitations of the wavemaker.
\begin{figure}[!ht]
	\begin{center}
		\includegraphics[trim=0.0cm 0cm 00cm 0cm, clip=true,width = 1\linewidth, height = 0.45\linewidth]{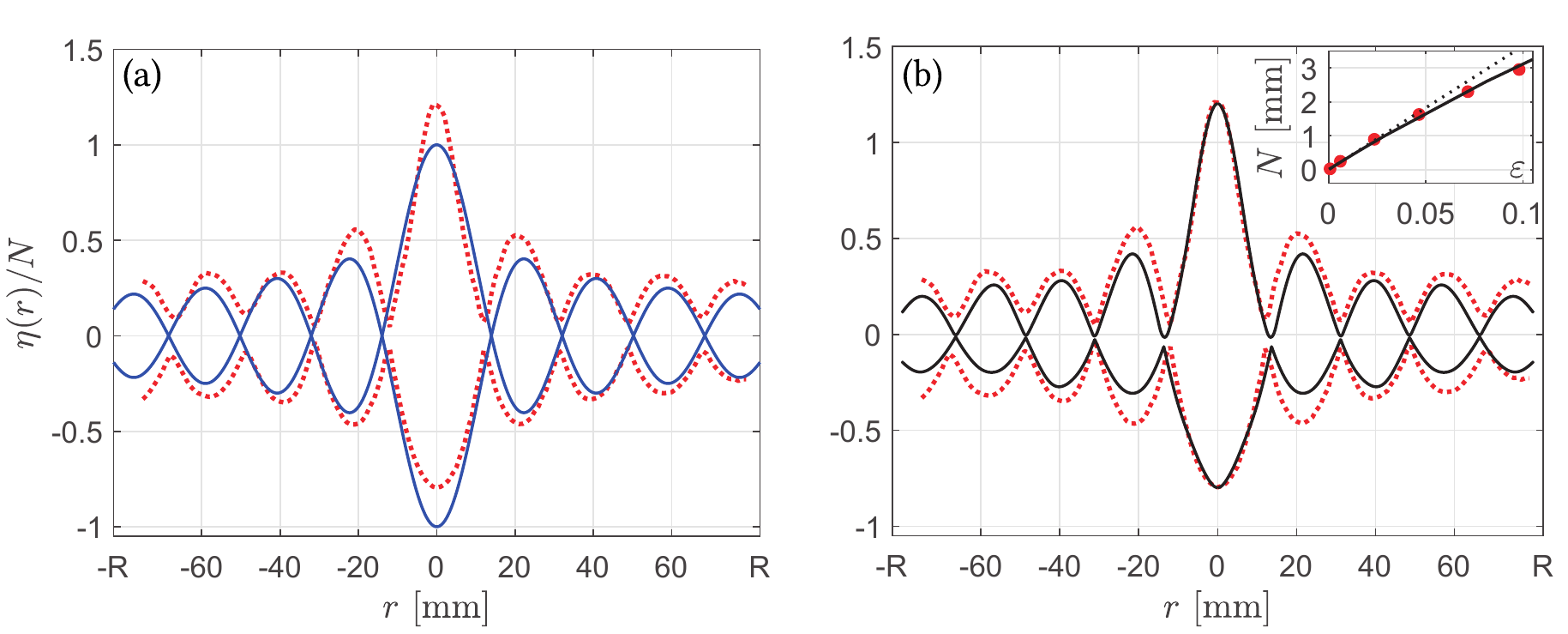}
		\caption{Dimensionless wave envelope along a basin diameter. All curves have been rescaled by the asymmetry coefficient $N = \flatfrac{(\emax - \emin)}{2}$. Red dotted line: experimental wave envelope showing both $\emax (r) >0$ and $\emin (r) < 0$ ($f = \SI{6.9}{\hertz}$, $a = \SI{0.38}{\centi\meter}$, $\veps = \num{0.36}$). (a) Blue solid line: Linear prediction from~\cite{Lamb-Book}. (b) Black solid line: Third-order nonlinear prediction computed numerically from~\cite{Mack1962}. Inset: $N$ vs. the nonlinear parameter $\veps$ for a fixed $f = \SI{6.9}{\hertz}$. Red dots correspond to data, black dotted line to the linear prediction, and black solid line to the nonlinear theory~\cite{Mack1962}.}
		\label{fig:Eta-max}
	\end{center}
\end{figure}

\section{Stationary spatial profile}\label{Profile}
We denote $\emax(r)$ the maximum of the wave elevation, $\eta (r,t)$, over time $t$ at position $r$, and the maximum central elevation $\emax\equiv\emax(0)$. In the same way, we define the quantities $\emin(r)$ and $\emin\equiv \emin(0)$ for the minimum of $\eta (r,t)$. We plot in Fig.~\ref{fig:Eta-max} the experimental wave envelope [i.e., $\emax (r)$ and $\emin (r)$] rescaled by $N = \flatfrac{(\emax-\emin)}{2}$ as a function of $r$ along a diameter (see red-dotted line in Fig.~\ref{fig:Eta-max}a-b). We superimposed in~Fig.~\ref{fig:Eta-max}a the prediction from the linear theory (blue solid line) when imposing two boundary conditions: (i) $\pdv*{\eta}{r}|_{r=0}= 0$ to ensure continuity at the basin center $r=0$, and (ii) $\eta (\pm R,t) = b \cos(2\pi f t)/2$ as the fluid must follow the ring oscillations at $r=\pm R$, and the fluid elevation $b$ may differ from the ring amplitude $a$. This leads to the envelope equation $\eta(r) = \emax J_0(kr)$ where $J_0$ is the Bessel function of the first kind and $k$ is computed from Eq.~\eqref{eq:disprelation} for fixed $f$, whereas $\emax$ is not needed thanks to the rescaling by $N$. We first notice that the linear theoretical profile in Fig.~\ref{fig:Eta-max}a does not show a zero slope at $r = \pm R$ indicating that the no penetrability condition used in Sect.~\ref{phase} is indeed debatable. Moreover, several differences are visible between the experiment and the linear model. First, the amplitude of the central deformation is measured to be asymmetric which is not captured by the linear theory. Second, the shift of the zeros suggests that the dispersion relation does not hold as is for nonlinear waves. Third, the linear prediction shows local minima with strictly zero vibration whereas the experiment shows nonzero minima of the envelope. These differences are significantly reduced when using a third-order nonlinear theory of axisymmetric gravity standing waves~\cite{Mack1962}. Indeed, Fig.~\ref{fig:Eta-max}b shows that the experimental and theoretical local minima occur at the same positions evidencing the relevance of using this nonlinear theory. The latter also confirms that the wave elevation at these nodes does not have to go to zero (although getting closer than in the experiment), in particular near the basin center. Indeed, the nonlinear theory predicts a slight horizontal oscillation of the locations of the zeros over a period so that the surface never goes flat and the water level in any location $r$ is nonzero at least for a fraction of time. Moreover, close to the focus, i.e., $r \to 0$, the up-and-down asymmetry (which is a classical signature of the nonlinearity~\cite{Stokes1847}) is well fitted by the nonlinear theory~\cite{Mack1962} (see black solid line). More precisely, the inset of Fig.~\ref{fig:Eta-max}b shows the asymmetry, $N = \flatfrac{(\emax - \emin)}{2}$, as a function of $\veps = \flatfrac{\emax}{\lambda}$. From the linear theory, one should have $\emax = - \emin$ leading to $N = \emax = \veps \lambda$ as displayed by the dotted line in the inset of Fig.~\ref{fig:Eta-max}b. The nonlinear theory computed numerically from Ref.~\cite{Mack1962} is also shown (solid line) and is found to be in good agreement with the experiments. Note that the departure between linear and nonlinear theories reaches 10\% for $\veps = 0.064$ which is close to the arbitrary criterion $\veps = 0.1$ used in Sect. III to distinguish the linear and nonlinear regimes. Finally, Fig.~\ref{fig:Eta-max} shows that the experimental profile near $r=\pm R$ does not satisfy the condition $\pdv*{\eta}{r}|_{r = \pm R} = 0$ as imposed to the system when computing its eigenmodes in \S \ref{phase}. This effect is experimentally confirmed for other forcing frequencies and could contribute to small departures between the resonances and eigenfrequencies observed on the phase diagram in Fig.~\ref{fig:PhaseDiagram}.

\section{Central deformation amplitude}\label{cda}
\begin{figure}[!htb]
	\begin{center}
		\includegraphics[trim=0cm 0cm 0cm 0cm, clip=true, width=0.95\linewidth]{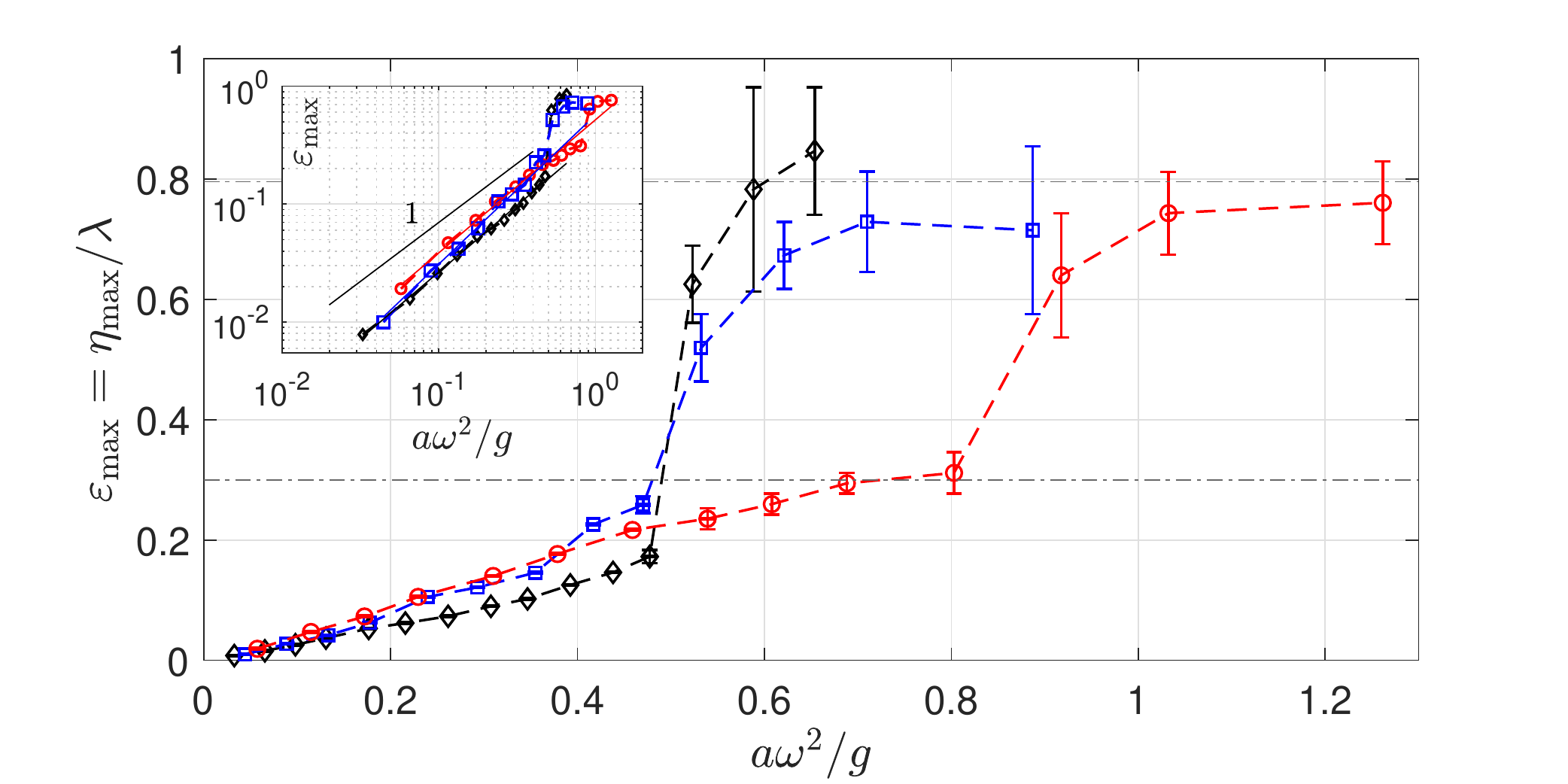} 
		\caption{Rescaled central maximum height, $\emax/\lambda$, as a function of the forcing acceleration, $a\omega^2/g$ (i.e., various forcing amplitudes), for three different forcing frequencies $f=$ ({\color{black}$\Diamond$}) $5.70$, ({\color{blue}$\square$}) $6.64$, and ({\color{red}$\circ$}) $\SI{7.55}{\hertz}$. Horizontal dot-dashed lines correspond to $\veps_\text{max} = 0.3$ and $5/(2\pi)$.
		Inset: Same in log-log scales. Solid line has a unit slope. Best fit has a 1.13 slope.}
		\label{fig:HauteurForcage}
	\end{center}
\end{figure}

The maximum elevation at the center ($r=0$) is now investigated. We measure it either by the capacitive probe for weak and moderate forcing amplitude, $a$, or by using a side camera (Basler $2048\times1536$~px$^2$, 120~fps) for higher~$a$. Error bars are the statistical average of data from a few similar jets, as the jet eruption often deviates from the vertical. Figure~\ref{fig:HauteurForcage} then shows the maximum height, $\emax$, reached by the fluid at the center rescaled by $\lambda$ as a function of the dimensionless forcing acceleration, $a\omega^2/g$, when varying the forcing amplitude $a$. The three different curves correspond to three different forcing frequencies $f\equiv \omega/(2\pi)$. In the linear and weakly nonlinear regimes ($\veps < 0.3$), $\emax/\lambda$ grows linearly with $a\omega^2/g$ regardless of $f$ as confirmed by the inset of Fig.~\ref{fig:HauteurForcage}. In the ejection regime ($\veps \gtrsim \num{0.5}$), the rescaled maximal height of the jet (not taking into account possible drop ejection) increases strongly with $a\omega^2/g$, then is found to saturate to a value denoted by $\veps_\text{sat} = \flatfrac{\esat}{\lambda}$, roughly independent of the acceleration, but depending on $f$. A model described below will explain this saturation. 

\section{Saturation model}
\begin{figure}[!htb]
	\begin{center}
		\includegraphics[trim=0cm 0cm 0cm 0cm, clip=true,width=0.7\linewidth]{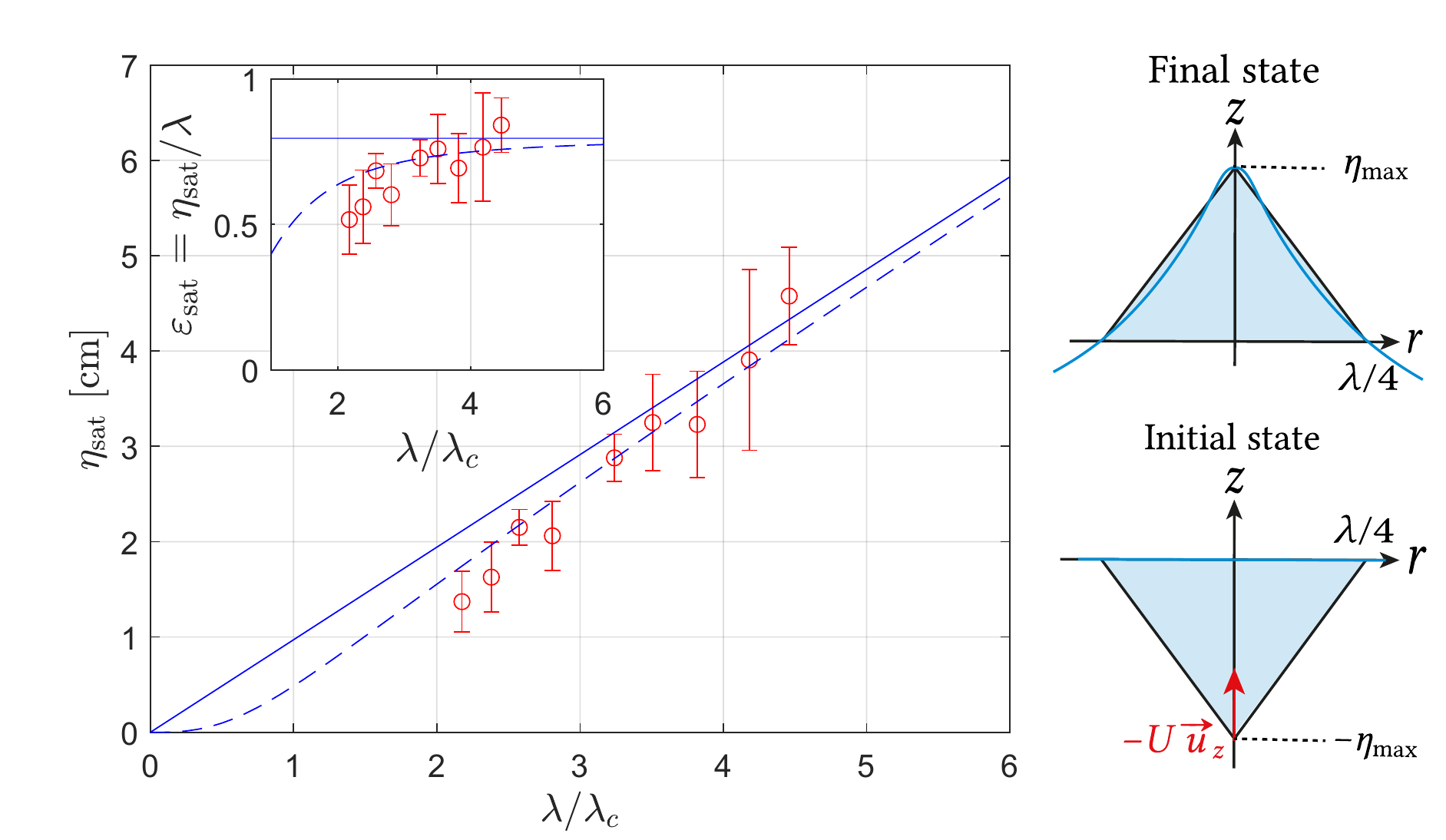} 
		\caption{Saturation wave height at the basin center, $\esat$, as a function of the rescaled wavelength, $\flatfrac{\lambda}{\lambda_c}$. ({\color{red}$\circ$}) experimental data for various forcing frequencies $f\in[5.5,8.5]$ Hz. (Blue solid line) prediction from the model of Eq.~\eqref{eq:esp-max} with $\alpha=\sqrt{2}$. (Blue dashed line) asymptotic trend of Eq.~\eqref{eq:esp-max} with $\alpha=\sqrt{2}$, i.e., $\esat^g= 5/(2\pi) \lambda$, valid in the pure gravity regime ($\lambda \gg \lambda_c$). Inset: Corresponding wave steepness at the basin center, $\veps_\text{sat} \equiv \flatfrac{\esat}{\lambda}$, vs. the dimensionless wavelength, $\flatfrac{\lambda}{\lambda_c}$. For large enough $\lambda$, the saturation steepness $\veps_\text{sat}$ tends to $\flatfrac{5}{(2\pi)}$. Same curves as in the main figure. Right: Schemes and quantities used in the model: (top) crude jet-shape approximation as a cone in its final state, (bottom) initial state with a flat surface and an upwards underwater-conical flow.} 		\label{fig:JetSaturation}
	\end{center}
\end{figure}

To explain the vertical saturation of the jet steepness, we approximate the jet surface at any time $t$, by a cone of height $\eta(t)$ and radius $\lambda/4$ as shown in the top-right hand Fig.~\ref{fig:JetSaturation} ($\flatfrac{\lambda}{2}$ is the natural diameter of the central deformation). First, we estimate the dominant forces of the problem. Given $\rho$, $\gamma$, $g$, $\mu=\SI{e-3}{\pascal\second}$ the dynamic viscosity of water, $L$ the typical size of the cone, and $v$ its typical vertical velocity, we compute the Weber (We), Reynolds (Re) and Bond (Bo) numbers. Taking $L = \emax \approx \SI{2}{\centi\meter}$ and $v = 2 \pi f \emax \approx \SI{0.8}{\meter\per\second}$, one obtains $\mathrm{We} = \rho v^2 L/\gamma \approx 350$, $\mathrm{Re} = \rho v L/\mu \approx 1.6\times10^{4}$, and $\mathrm{Bo} = \rho g L^2/\gamma \approx 100$. The Weber number is defined as the ratio between the inertial and surface tension forces and shows that inertia dominates surface tension effects. The Reynolds number (comparing inertia forces to viscous ones) shows that viscosity can be neglected here. The Bond number shows that gravitational forces are two orders of magnitude larger than surface tension ones. Since $\mathrm{We}$, $\mathrm{Re}$, and $\mathrm{Bo} \gg 1$, viscous and surface tension forces can be neglected. The following model thus takes into account only inertial and gravitational forces through kinetic and potential energies. We now consider the energy balance between a final state where the jet of height $\emax$ has a finite conical volume $V$, a positive potential energy and is motionless, and an initial state where the surface is flat and the same volume of the fluid is enclosed within a downwards cone of same dimensions located under the free surface (see bottom-right hand Fig.~\ref{fig:JetSaturation}). Leaving out the $t$ notation for clarity, the cone edge equation thus reads
\begin{equation}
	z(r) = \pm \emax \mp \frac{4\emax}{\lambda} r\, , \qq{i.e.,} r(z) = (\emax \mp z) \frac{\lambda}{4 \emax} \, ,
	\label{eq:Edgecone}
\end{equation}
where the upper (resp. lower) signs describe the top (resp. bottom) cone and $z$ is the vertical coordinate. The volume of such a cone reads $V = \flatfrac{\pi \lambda^2 \emax}{48}$. The fluid velocity in the bottom cone is unknown, but we keep only its vertical component $u(z)$ and approximate it by a linear dependence on $z$ between $u(0)=0$ and $u(-\emax) = U$ (upward velocity at its lowest depth), as proposed and proven sufficient in Ref. \cite{Che2018}. This yields

\begin{equation}
	u(z) = - \frac{z}{\emax}U\, , \ \ {\rm for}\ \ z\in[-\emax,0]\, ,
	\label{eq:speed}
\end{equation}

Then, we can express in a general way the kinetic energy $E_k = \int_{-\emax}^0 \rho \pi r(z)^2 u(z)^2 \dd z/2$ and the potential energy $E_p = \int_{-\emax}^0 \rho \pi r(z)^2g z \dd z$ of the bottom cone. Substituting Eq.~\eqref{eq:Edgecone} and~\eqref{eq:speed} in these expressions, we end up with $E_k = \flatfrac{\rho V U^2}{20}$ and $E_p = - \flatfrac{\rho g \emax V}{4}$. For the final state (top cone), $E_k = 0$ as $U = 0$ and $E_p =+\flatfrac{\rho g \emax V}{4}$. Neglecting as justified above, viscous dissipation and surface tension, the conservation of energy between the final state and the initial state yields $\emax = \flatfrac{10g}{U^2}$. To find the dependence of $\emax$ on the wavelength $\lambda$, we assume that the dispersion relation for linear waves of Eq.~\eqref{eq:disprelation} makes a first approximation even for these nonlinear and nonsinusoidal deformations. Taking $U$ of order $\alpha \omega \emax$ with $\alpha$ a fitting parameter, this yields
\begin{equation}
	\emax = \frac{5}{\alpha^2 \pi} \frac{\lambda}{[1+(\lambda_c/\lambda)^2]}\, ,
	\label{eq:esp-max}
\end{equation}
where $\lambda_c/(2 \pi) \equiv \sqrt{\flatfrac{\gamma}{\rho g}} = \SI{1.9}{\milli\meter}$ is the capillary length separating the capillary ($\lambda \ll \lambda_c$) and gravity ($\lambda \gg \lambda_c$) wave regimes, the gravity-capillary regime occurring in between.
Note that, in the capillary-wave regime, the surface tension cannot be neglected anymore and a different model must be applied when $\mathrm{Bo} \lesssim 1$, i.e., $\lambda \lesssim \SI{2}{\centi\meter}$. We plot $\emax(\lambda)$ from Eq.~\eqref{eq:esp-max} in Fig.~\ref{fig:JetSaturation} together with the experimental maximum heights reached at the basin center for our gravity-capillary range ($\lambda \in [2.5, 4.5] \, \si{\centi\meter}$). As in \S \ref{cda}, error bars come from the statistical average of data from a few similar jets. Figure~\ref{fig:JetSaturation} then shows that our model well captures the experimental saturation heights using $\alpha = \sqrt{2}$. $\alpha > 1$ means that the linear approximation $U=\omega \emax$ underestimates the cone velocity at saturation and that nonlinear effects tend to increase it. However, the saturation of the jet height indicates that beyond a certain forcing amplitude the energy injected in the system does not contribute to the central jet velocity, which should be directly converted into potential energy, but is mainly dissipated by the meniscus at the ring boundaries and/or transferred within the fluid bulk in the form of flows. Equation~\eqref{eq:esp-max} with $\alpha=\sqrt{2}$ then tends asymptotically in the pure gravity regime to $\esat^g=\flatfrac{5 \lambda}{(2\pi)}$, corresponding thus to a saturation of the wave steepness towards $\veps^g_\text{sat}\equiv \esat^g/\lambda=\flatfrac{5}{(2\pi)}$. The inset indeed shows that this predicted saturation of the wave steepness occurs experimentally for large enough wavelengths. Beyond the good agreement, the comparison is experimentally limited by viscous dissipation and mechanical limitation of the shaker for smaller wavelengths, and by the system size for larger wavelengths.

Finally, note that for existing large-scale axisymmetric tanks (from 1.6 m to 25 m diameters) surrounded by several wavemakers, driven in unison \cite{Minoura2011,Maeda2004,McAllister-JFM2022}, Eq.~\eqref{eq:esp-max} would give an upper bound of the height of the central wave crest, which would be either larger than the size of the tank building, or not reachable due to wavemaker limitations. Furthermore, Eq.~\eqref{eq:esp-max} should not be confused with another limit commonly discussed in the literature concerning the stability of a periodic sharp-crested wave. Indeed, for unidimensional progressive gravity waves of finite amplitude, this limiting angle $\beta$ of the crest was first derived by Stokes to be \SI{120}{\degree} assuming a steady crest profile~\cite{Stokes1847}, and to be \SI{90}{\degree} for the standing wave case~\cite{Penney1952}, the latter value being confirmed experimentally~\cite{Taylor1953}. For axisymmetric standing gravity waves of finite amplitude, a limiting angle of \SI{109.47}{\degree} was derived analytically~\cite{Mack1962} and tested experimentally~\cite{Fultz1963,McAllister-JFM2022}, corresponding thus to a rescaled maximum height $\emax/\lambda = 1/(4 \tan{\beta/2}) = 0.25$. As expected, this value is much smaller than $0.8$ as it corresponds to a stability limit (and not to a maximum height). 

\section{Conclusion}We reported on the experimental study of the focusing of axisymmetric gravity-capillary waves generated by a vertically vibrating ring partially immersed in a fluid. Different regimes of standing waves are observed at the basin center depending on the forcing parameters: linear, nonlinear, and ejection regimes. For weak forcing, and close to a natural frequency of the circular basin predicted by the linear theory~\cite{Lamb-Book}, the standing waves display a resonant response. For stronger forcing, we observed that the spatial profile of standing waves breaks the up-down symmetry, and exhibits nonzero local minima, which are both well taken into account by a nonlinear theory of axisymmetric standing waves up to third order in amplitude~\cite{Mack1962}. Finally, for an even stronger forcing, we observed a jet together with possible drop ejections. The maximum elevation reached experimentally by the wave at the center of the basin is found to saturate at $\esat$, even for stronger forcing amplitudes. For gravity waves, $\esat$ increases linearly with the wavelength, due to the saturation of its steepness to $5/(2\pi)$. This maximum wave height is well captured using a crude model, based on an energy balance with strong hypotheses concerning the forces and the shape of the jet. This is a first step towards a more elaborated one. 
In the future, we will address the origin of the jet. Does it arise out of a deep depression of the free surface leading to the collapse of this cavity coupled to a singularity, or/and the collapse of a bubble entrapped underneath~\cite{Zeff2000,Josserand2017}? The dynamical properties of the focusing will be also investigated by tracking the propagation of axisymmetric gravity-capillary propagating waves converging towards the center to explore open questions, such as which mechanisms drive their central interaction, and how the power injected by the ring is dissipated at the central singularity?
\begin{acknowledgments}
We thank G. Michel and M. Roch{\'e} for fruitful discussions. We thank A. Di Palma, and Y. Le Goas for technical help. This work is supported by the French National Research Agency (ANR DYSTURB project No. ANR-17-CE30-0004, ANR SOGOOD project No. ANR-21-CE30-0061-04), and by the Simons Foundation MPS No. 651463-Wave Turbulence (USA).
\end{acknowledgments}

\appendix

\section{Role of the immersed depth of the ring}\label{AppendA}
\begin{figure}[!ht]
	\centering
	\includegraphics[width = 7cm]{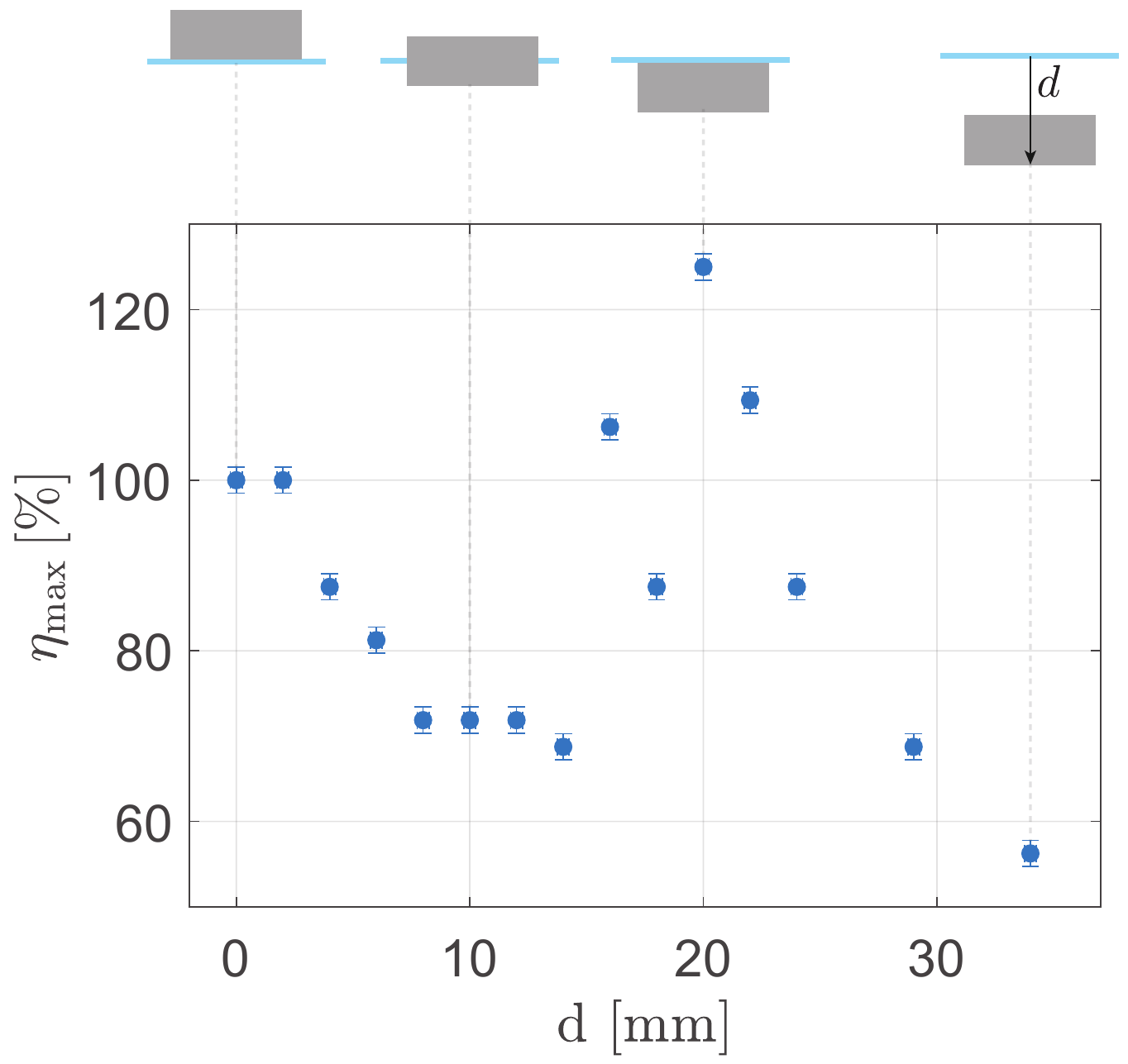}
	\caption{(Top) Sectional schemes defining $d$ as the ring distance to the water surface at rest (horizontal blue line). The ring thickness (gray area) is 20 mm. (Bottom) Maximum water elevation, $\eta_\text{max}$, measured at the center when increasing $d$ for fixed sinusoidal forcing parameters ($f = 6.9$~Hz and $a = 0.13$~cm), and plotted as a percentage of the initial value.}

	\label{fig:InfluenceProfondeur}
\end{figure}

We quantify the influence of the ring immersion into the fluid on the wave properties. We measure the maximum water elevation, $\eta_\text{max}$, at the center, for different relative positions $d$ of the ring to the water surface at rest, all other things being equal (especially the forcing parameters $f$ and $a$). The experimental data are displayed in Fig.~\ref{fig:InfluenceProfondeur} and show that $\eta_\text{max}$ depends on the fluid volume moved by the ring. Indeed, $\eta_\text{max}$ is maximum when the ring is fully immersed into the fluid and flushing to the surface ($d=20$~mm). The water volume moved by the ring oscillation, when the ring is closer to the free surface, is thus combined with the meniscus movement (same for $d\approx 0$~mm). On the other hand, when the ring is partially immersed ($d\approx 10$~mm), $\eta_\text{max}$ reaches a plateau where only the meniscus effect plays a role in the wave forcing. $\eta_\text{max}$ decreases as expected when the ring plunges deeper and deeper ($d$ larger than the ring thickness 20~mm) and the two forcing effects disappear. Beyond this dependence of the wave amplitude on $d$, we have furthermore verified that the initial immersed depth of the ring does not impact other wave properties.
\section{Experiments with a different ring size}\label{AppendB}
\begin{figure}[!htb]
	\centering
	\includegraphics[trim=0cm 0cm 0cm 0cm, clip=true,width = 0.5\linewidth, height= 0.36\linewidth]{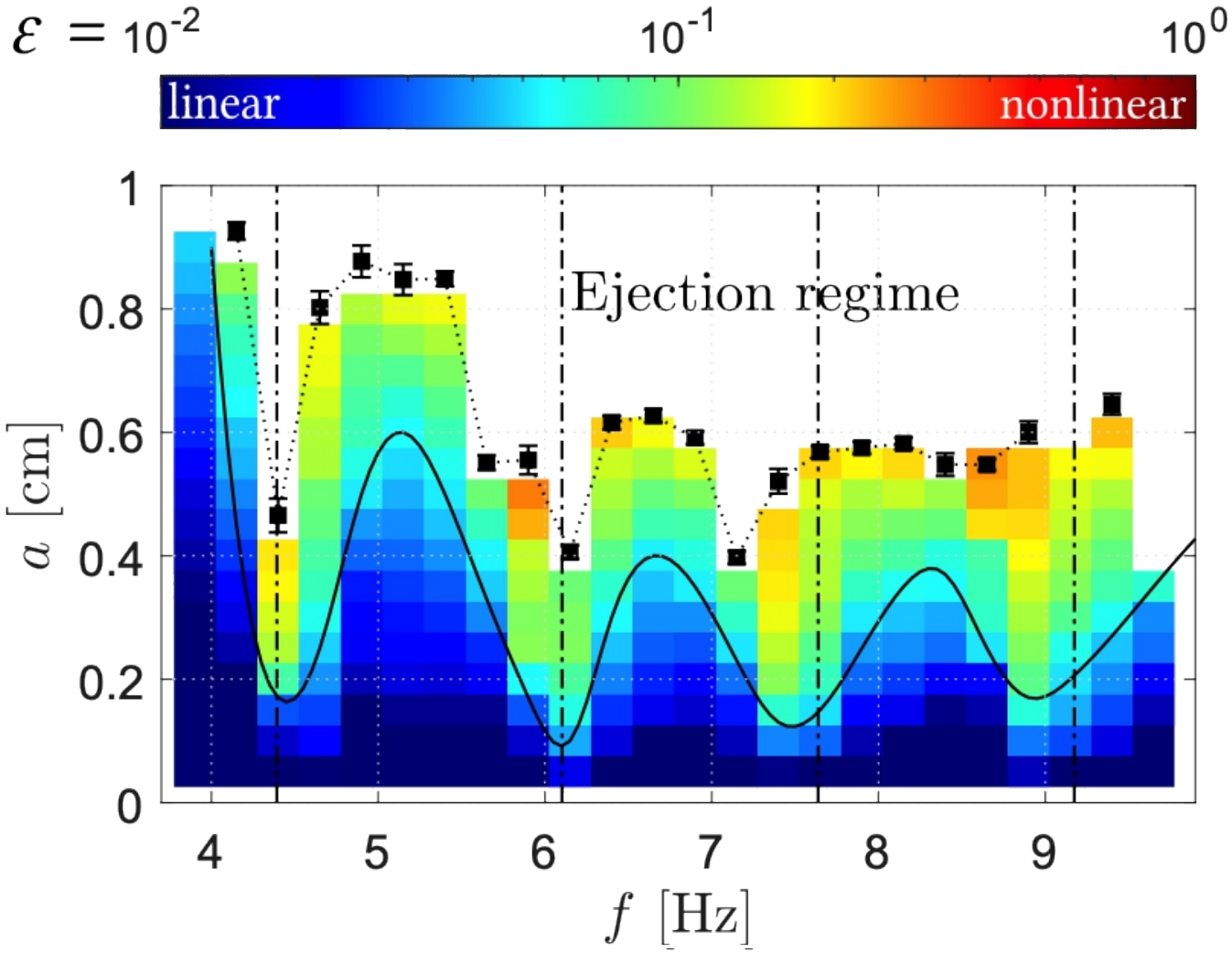}
	\caption{Wave steepness, $\veps= \flatfrac{\eta_\text{max}}{\lambda}$, of the central deformation as a function of the forcing amplitude $a$ and frequency $f$. Logscale colorbar. Vertical lines: circular basin eigenfrequencies $f_n$ from $J_0'(k_nR) = 0$ (here $R = \SI{5.1}{\centi\meter}$, compared to Fig.~\ref{fig:PhaseDiagram}) and $f_n$ and $k_n$ are related by Eq.~\eqref{eq:disprelation}. Solid line corresponds to the same value of $\veps \simeq \num{0.1}$ as a function of $f$. ($\blacksquare$): maximum amplitude of the central deformation before ejection vs. $f$.}
	\label{fig:PhaseDiagram_Small}
\end{figure}

We perform the same experiments as in the main text, but with a ring of different radius. We now use a ring radius of $R = \SI{5.1}{\centi\meter}$ instead of $\SI{8.25}{\centi\meter}$ as in the main text. Figure~\ref{fig:PhaseDiagram_Small} shows the corresponding phase diagram as a function of the control parameters. The resonance frequencies with this smaller ring are $f = 4.37$, 6.07, 7.59, and 9.11~\si{\hertz}, which differ from those with the larger ring in Fig.~\ref{fig:PhaseDiagram} (i.e., 5.7, 6.66, 7.59, 8.53, 9.48~\si{\hertz}), but correspond to the circular basin eigenmodes. It thus confirms that the system resonance frequencies are indeed given by the ring eigenfrequencies.


\begin{thebibliography}{99}
\bibitem{note}Note that the Bessel function minima are not exactly periodic, in particular for the first two ones (departures of 0.8\% and 0.2\%, respectively), but are approximately periodic for the following ones (departures less than 0.01\%).
\end{thebibliography}
\end{document}